\documentclass[preprint]{revtex4-1}

\usepackage{graphicx}
\usepackage{amssymb}
\usepackage{dcolumn}
\usepackage{bm}

\begin{document}

\bibliographystyle{unsrt}

\title {Propagation of ultra-short solitons in stochastic Maxwell's equations}

\author{Levent Kurt}
\affiliation{Division of Mathematics and Natural Sciences\\                                        
                Queens College \\  
                City University of New York \\                              
                Queens, NY}
\author{Tobias Sch{\"a}fer}
\affiliation{Department of Mathematics\\                                  
          College of Staten Island \\                                           
          City University of New York \\                                        
          Staten Island, NY}
\email{tobias@math.csi.cuny.edu}

\begin{abstract}
  We study the propagation of ultra-short short solitons in a cubic
  nonlinear medium modeled by nonlinear Maxwell's equations with
  stochastic variations of media. We consider three cases: variations of (a) the 
  dispersion, (b) the phase velocity, (c) the nonlinear coefficient. 
  Using a modified
  multi-scale expansion for stochastic systems, we derive new
  stochastic generalizations of the short pulse equation that
  approximate the solutions of stochastic nonlinear Maxwell's equations.
  Numerical simulations show that soliton solutions of the short pulse
  equation propagate stably in stochastic nonlinear Maxwell's equations
  and that the generalized stochastic short pulse equations approximate
  the solutions to the stochastic Maxwell's equations over the distances under
  consideration. This holds for both a pathwise comparison of the stochastic equations
as well as for a comparison of the resulting probability densities.
\end{abstract}


\maketitle

\section{Introduction}

Many integrable partial differential equations arise as asymptotic
expansions of nonlinear wave equations. In the particular context of
ultra-short pulses propagating in cubic nonlinear media, the basic
model equation is given by Maxwell's wave equation with dispersion and
nonlinearity coming from the material's response to the excitation by
the electric field \cite{boyd:1992}. When the pulse width is large in
comparison to the carrier wavelength, an asymptotic expansion of the
nonlinear wave equation yields the {\em cubic nonlinear
  Schr{\"o}dinger equation} (NLSE) \cite{newell-moloney:1992}. The
NLSE is integrable \cite{zakharov-shabat:1971} and has bright soliton
solutions that have been studied intensively over the last four
decades, not only because of their mathematical beauty, but also
because of a variety of possible applications in modern optical
technology \cite{hasegawa-kodama:1995}. Recent experimental progress
in the field of ultra-short pulses \cite{karasawa-nakamura-etal:2001}
has led to increased interest in models that go beyond the NLSE
description: As the length of the pulse shortens, the basic assumption
of scale separation between envelope and carrier wave is not satisfied
anymore \cite{rothenberg:1992} and additional correction terms have to
be added \cite{agrawal:2007}. On the other hand, it is possible to
introduce a different scaling \cite{alterman-rauch:2000} in the
asymptotic analysis of the nonlinear wave equation that is appropriate
for ultra-short pulses. In this way, it is possible to derive a
different approximation of the solution of Maxwell's equation, the
{\em short pulse equation} (SPE) \cite{schaefer-wayne:2004}. Although
derived under entirely different assumptions in comparison to the NLSE, the SPE is
integrable \cite{sakovich-sakovich:2005} as well and was shown to
possess bright solitons that may be as short as three cycles
\cite{sakovich-sakovich:2006}. Consequently, over the last years, the
mathematical structure of the SPE was subject to intensive research
\cite{brunelli:2005,sakovich-sakovich:2007,matsuno:2007,victor-thomas-etal:2007,manukian-costanzino-etal:2009,pelinovsky-sakovich:2010}.

When a solitary wave solution is taken as the initial condition in the
SPE, it persists for arbitrary propagation distances. As initial
condition of the original nonlinear wave equation, however, it is expected to
change its shape due to the fact that the SPE is only an approximation
to the nonlinear wave equation. If $\epsilon$ is the expansion factor
in the multiple-scale asymptotic analysis, we would expect to see a
noticeable deviation only after propagation distances of the order of
$1/\epsilon$ \cite{holmes:1995}. We show at the end of the following
section numerically by propagating the soliton to a distance of order
$1/\epsilon^2$ that this change happens indeed very slowly, such that
we can say that the SPE solitons persist in Maxwell's equations. The 
slow change of the soliton solution is governed by higher-order effects that
appear as additional terms in the multi-scale expansion and that can be studied
systematically \cite{kurt-chung-etal:2012}. In the present paper, since we are focusing 
on the leading order impact of stochastic perturbations of the parameters in the 
equations, we neglect these higher-order deterministic effects.
Incorporating deterministic higher-order effects in the presence of
random perturbations lies beyond the scope of this paper and will be
studied in the future.

Such random fluctuations are widely present in nature and models in which stochasticity is taken into account are more realistic than deterministic models. One important question that arises in the context
of integrable systems is whether soliton solutions persist when perturbed stochastically: NLSE solitons broaden under the influence of stochastic variations of the dispersion coefficient \cite{abdullaev-bronski-etal:2000} . For the NLSE case, however, this broadening is sufficiently slow such that the soliton still persists over long distances, provided that the stochastic perturbations are sufficiently small. One result of this paper is that SPE solitons show similar
behavior.

When studying the impact of stochastic variations of the media, one faces the
problem of {\em coarse-graining} noise, namely the question of how microscopic variations of the parameters in the nonlinear wave
equation relate to variations of the parameters in the SPE or the
NLSE. In section \ref{sec:stochastic_SPE}, based on previous work
\cite{schaefer-moore-etal:2002,schaefer-moore:2007}, we show how this
coarse-graining of noise can be done explicitly in the multi-scale
expansion that leads from the nonlinear wave equation to the SPE: We
introduce randomness in the linear dispersive part of the
susceptibility of the media and obtain a stochastic dispersion
coefficient in the SPE, yielding a stochastic generalization of the
short pulse equation. Similarly, stochasticity can be introduced in the phase
velocity and and in the nonlinearity. The resulting generalized stochastic short pulse equations
constitute the main result of this work.

It remains to be shown that the solutions to the stochastic short pulse equations stay close
to the solutions of the original stochastic nonlinear wave equation. Even in the deterministic
context this is very challenging and, in fact, this problem has only be solved recently \cite{pelinovsky-schneider:2011}. Therefore, in section \ref{sec:numerics} 
we present a numerical comparison of the solutions to the nonlinear wave equation with randomness and 
the corresponding stochastic short pulse equations. We show that,
over the distances under consideration, SPE solitons also persist in
the stochastic short pulse equations and in the stochastic nonlinear
wave equation. Moreover, extensive numerical simulations show that the
SPE model can be used to predict the likelihood of large deviations from
the deterministic solution when we compare particular stochastic 
paths and as well as the corresponding probability distributions.

\section{Ultra-short solitons in deterministic Maxwell's equations}

In this section we briefly review the setting of the short pulse
equation and its derivation from Maxwell's equations. We first
consider the following deterministic one-dimensional wave equation
describing pulse propagation in bulk silica \cite{schaefer-wayne:2004}
\begin{equation} \label{maxwell_1d}
u_{xx}-u_{tt} = \chi_0 u +\chi_3(u^3)_{tt}\;.
\end{equation}
Here, $x$ is our evolution variable and we assume that we are given
initial conditions $u(x=0,t)$ and $u_x(x=0,t)$. In this section we
assume the coefficients $\chi_0$ and $\chi_3$ to be fixed, in the next
section we will modify the problem by allowing the coefficients $\chi_0$ and $\chi_3$ to vary
stochastically. Above equation can be directly derived from Maxwell's
equations under the assumption that the response of the medium is
instantaneous, otherwise a retarded response of the material will lead
to convolution integrals in the polarization which will yield a
nonlocal wave equation \cite{chung-schaefer:2007}. We also assume that
the Fourier transform $\hat u$ is zero for very low wavelengths. This
accounts for the fact that such waves cannot propagate in the optical
medium.

As for many wave equations, we can use asymptotic expansions in order
to construct approximate solutions to (\ref{maxwell_1d}). The short
pulse equation (SPE) can be derived from (\ref{maxwell_1d}) using a
multiple-scale expansion \cite{schaefer-wayne:2004} or an equivalent
method like the renormalization group method
\cite{chung-jones-etal:2005}: A multi-scale expansion of the form
\begin{equation}{\label{uExpansion}}
u(x,t)=\epsilon A_0(\phi,x_1,x_2,...)+\epsilon^2 A_1(\phi,x_1,x_2,...)+...
\end{equation}
with
\begin{equation}{\label{variablesExpansion}}
\phi=\frac{t-x}{\epsilon}, \qquad x_n=\epsilon^n x\,
\end{equation}
yields as leading order 
\begin{equation} \label{spe_sw}
-2 \partial_{\phi} \partial_{x_1} A_0 = \chi_0 A_0 + \chi_3 \partial^2_{\phi} A_0^3\,. 
\end{equation}
This form of the short pulse equation can be transformed to the standard form
\begin{equation} \label{spe}
U_{XT}=U+\frac{1}{6}U^3_{XX}
\end{equation}
through the transformation given by
\begin{equation} \label{transformation}
A_0(\phi,x_1)=\sqrt{\frac{\chi_0}{6\chi_3}}U(X,T), \qquad X=-\phi, \qquad T=\frac{\chi_0}{2}x_1\,.
\end{equation}
Note that we change here from the physical variables $(x_1,\phi)$
(which are rescaled and shifted with respect to the variables $(x,t)$
of the nonlinear wave equation) to the mathematical variables
$(T,X)$. In all simulation results later, we will always use the
physical variables $(x,t)$ of the original nonlinear wave
equation. 

To our knowledge, it is unknown whether the original wave
equation (\ref{maxwell_1d}) is integrable. Sakovich and Sakovich, however, have
shown that the SPE (\ref{spe}) is integrable and, moreover, admits
one-soliton solutions \cite{sakovich-sakovich:2006} that can be written
as
\begin{eqnarray}
U & = & 4mn\frac{m\sin\psi\sinh\phi+n\cos\psi\cosh\phi}{m^2\sin^2\psi+n^2\cosh^2\phi} \nonumber \\ 
X & = & Y+2mn\frac{m\sin2\psi - n\sinh2\phi}{m^2\sin^2\psi+n^2\cosh^2\phi} 
\label{sakovich_soliton}
\end{eqnarray}
with
\begin{equation}
\phi=m(Y+T),\qquad \psi=n(Y-T), \qquad n=\sqrt{(1-m^2)}\,. 
\end{equation}
This family of solutions depends on the parameter $m$, and the
condition for these solitary wave solutions being non-singular was
found to be
\begin{equation}
m<m_{cr}=\sin\frac{\pi}{8} \approx 0.383.
\end{equation}
The parameter $m$ determines the shortness of the pulse. As $m$
reaches its critical value, the shortest pulse of this family we can construct is
approximately a three cycle pulse. When $m$ is small (for instance,
$m=0.01$), solitary wave solution of the SPE can be approximated by
\begin{equation} \label{nls_like_soliton}
U \approx 4m\,\cos(X-T){\mathrm{sech}}(m(X+T))
\end{equation}
such that the pulse is similar to a bright soliton solution of the
cubic nonlinear Schr{\"o}dinger equation. In the present work, we focus on the
behavior of ultra-short solutions. In our simulations, we take $m=0.3$.

With these soliton solutions of the SPE at hand, the question arises
whether they also show stable propagation when taken as initial
conditions in the original problem (\ref{maxwell_1d}). First, it was
shown for the linear case ($\chi_3=0$) that the solution of the SPE
stays close to the solution of the original problem
\cite{chung-jones-etal:2005}. For the nonlinear case, the resulting 
proof is much more difficult and a solution was found only recently \cite{pelinovsky-schneider:2011}.

\begin{figure}[htb]
\centering
\scalebox{0.4}{\includegraphics{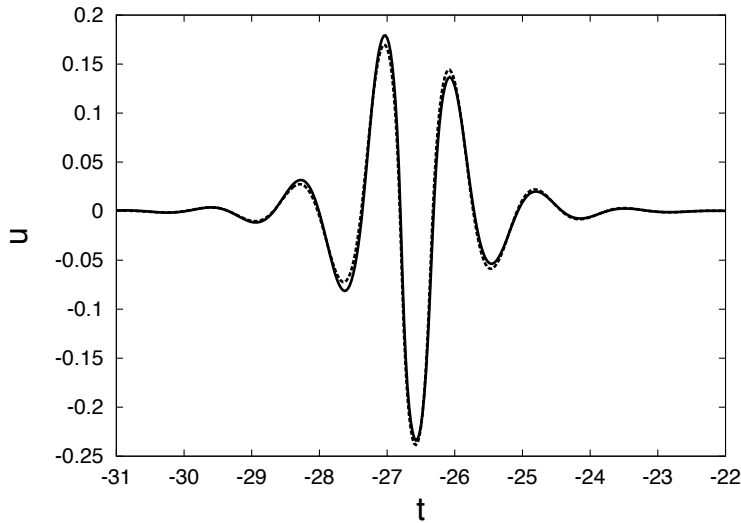}}
\hfill
\caption{Profile of the SPE soliton in the nonlinear wave equation (\ref{maxwell_1d}). The solid
line shows the snapshot of the soliton solution as a result of its propagation in Maxwell's nonlinear wave equation.
The dashed line is the corresponding solution to the short pulse equation with the same initial condition. The small deviations are due to higher-order deterministic corrections.}
\label{fig:comparison}
\end{figure}

Figure \ref{fig:comparison} shows a numerical solution of the nonlinear wave
equation (\ref{maxwell_1d}) with the SPE soliton solution taken as
initial conditions. In the simulation, we chose the parameter $m$ to
be $0.3$ and $\epsilon=0.2$. This snapshot of the soliton profile is compared to the
analytical solution of the SPE given by (\ref{sakovich_soliton}). The
soliton shows stable propagation in the nonlinear wave equation - note
that the evolution distance in $x$ is
$25.625\approx{\mathcal{O}}(1/\epsilon^2)$, which is significantly
further than we would expect for a leading order multi-scale
expansion. The reason for this can be found when considering addtional
higher-order terms in the multi-scale expansion \cite{kurt-chung-etal:2012}. 
In this work, however, we neglect such high-order contributions.

\begin{figure}[htb]
\centering
\scalebox{0.4}{\includegraphics{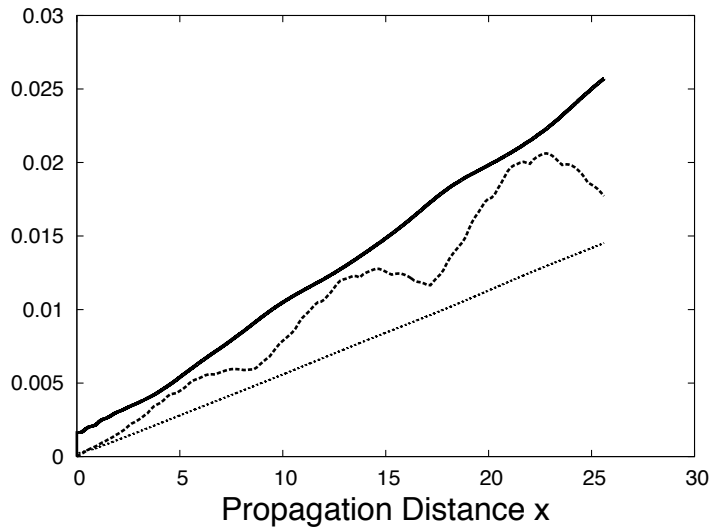}}
\hfill
\caption{Evolution of the deviations of the SPE soliton when
  propagating in the nonlinear wave equation (\ref{maxwell_1d}). The
  solid line shows the evolution of the $L^1$-norm given by
  $\|u-u_s\|_{L^1}$, the dashed line the evolution of the
  $L^{\infty}$-norm, and the dotted line the evolution of the
  $L^2$-norm.}
\label{fig:NormDifference}
\end{figure}

In general, we expect the difference between the solution $u$ of the
nonlinear wave equation and the soliton $u_s$ to grow with the
propagation distance $x$.  We can quantify this growth by a numerical
comparison. In figure \ref{fig:NormDifference} we plot the evolution
of several norms of $u-u_s$, in particular the $L^1$, $L^2$ and
$L^{\infty}$-norms defined by
\begin{equation}
\|f\|_{L^k}=\left(\int |f(t)|^k\,dt\right)^{1/k}, \qquad 
\|f\|_{L^{\infty}}=\max(|f(t)|)\,.
\end{equation}
%

Whereas the $L^1$ and the $L^2$-norms display monotonic growth, the $L^{\infty}$-norm
shows oscillations that are due to the fact that this norm is very sensitive to shifts in time
as it compares the solutions pointwise.

\section{The stochastic short pulse equation} \label{sec:stochastic_SPE}

Optical soliton propagation in fibers in the presence of a stochastic perturbation has been studied intensively in the context of the NLSE model \cite{abdullaev-bronski-etal:2000}. The sources of randomness in optical fibers may vary. For example, the phase of the wave may fluctuate due to the presence of the randomness. One major source of medium-related stochastic phase fluctuations are the inhomogeneities in a fiber's core or the fluctuations in the linear refractive index of the core. Stimulated Brillouin scattering, stimulated Raman scattering, and inhomogeneities of the medium are the possible sources of the phase fluctuations \cite{hart-judy-roy-beletic:1998,sulem-sulem:1999}.
The stochasticity may come from the nonlinearity of the medium as well. The dynamical effect of the noise added by the stochastic nature of the nonlinearity is much smaller than the noise due to the inhomogeneities \cite{abdullaev-hensen-bischoff-sorensen:1998}. Nevertheless, there may be other sources of randomness playing a role, and they may originate from the other parts of the system such as the inherent power fluctuation in lasers used as input pumps. Apart from these, quantum phase fluctuations are also sources of phase noise in optical fibers although they are practically negligible \cite{agrawal:2007}. In many applications, Langevin noise is used to study the fluctuations in the system \cite{boyd:1992}.

In the present work, we do not model the precise microscopic origins of the randomness, but
rather assume that they lead to small stochastic variations of the coefficients of the equations. 
This approach is common, one particular case studied in the past is the NLSE with a linear multiplicative stochastic term corresponding to stochastic variations
of the dispersion. In analogy, it is natural to ask how the ultra-short solitons of the deterministic SPE propagate in a stochastic environment.

Therefore, the simplest corresponding stochastic variations can be introduced to the wave equation (\ref{maxwell_1d}) by making the coefficient $\chi_0$ representing the linear susceptibility stochastic. Setting
\begin{equation} \label{stochastic_chi}
\chi_0 = \bar\chi_0 + \chi_{\mathrm{rand}}(x), \qquad \chi_{\mathrm{rand}}(x)=\nu\xi(x)\,,
\end{equation}
leads to the stochastic wave equation such that
\begin{equation} \label{maxwell_stochastic1}
u_{xx}-u_{tt} = \left[\bar\chi_0 +\nu\xi(x)\right] u +\chi_3(u^3)_{tt}\;,
\end{equation}
where $\xi$ represents white noise, hence $\langle
\xi(x)\xi(x')\rangle = \delta(x-x')$, and the coefficient $\nu$
measures the strength of the noise. Although mathematically
convenient, we note that this is only an approximation, as in reality,
the noise will be short-correlated rather than delta-correlated. In the
numerical simulations we also will assume that the coefficient of the
dispersive term $\chi_0$ will always be larger than zero which
correctly represents the underlying physical system.

In order to derive a stochastic generalization of the short pulse
equation, we introduce the same multi-scale expansion as in the
deterministic case of the form
\begin{equation}                                                                
u(x,t)=\epsilon M_0(\phi,x_0,x_1,x_2,...)+\epsilon^2 M_1(\phi,x_0,x_1,x_2,...)+...      
\end{equation}
with again rescaled variables
\begin{equation}                                                                
\phi=\frac{t-x}{\epsilon}, \qquad x_n=\epsilon^n x.                             
\end{equation}
As in the deterministic case, we solve now the equation order by
order. As terms of the order $1/\epsilon$ cancel, we obtain at
${\mathcal{O}}(1)$ the equation $M_{0\phi x_0}=0$ such that the
leading order $M_0$ is independent of the fast variable $x_0$. The
equation for $M_1$ is given by a combination of terms and can be
written as:
\begin{equation}\label{equation_for_M1}                                 
- 2(M_1)_{x_0 \phi} = 2(M_0)_{x_1\phi} 
+ ( \bar \chi_0 + \nu \xi(x)) M_0 + \chi_3 (M_0 ^3)_{\phi\phi} \,.          
\end{equation}
In the deterministic case, if $\nu=0$, the solvability condition of
this equation yields the short pulse equation, and therefore the
entire r.h.s. of (\ref{equation_for_M1}) will be zero. For the
stochastic case, the standard multi-scale argument has to be slightly
modified due to the presence of the noise term. If we integrate
(\ref{equation_for_M1}) with respect to the fast variable $x_0$ from
$x_0=0$ to $x_0=1$, the noise term $\xi(x)$ yields a random number
$\zeta_0$ given by $\zeta_0=\int_0^1\xi(x)dx$. As $\xi$ is white
noise, $\zeta_0$ is normally distributed with mean zero and variance
one. On the slow scale $x_1$, the points 0 and 1 correspond to $0$ and
$\epsilon$ and $\zeta_0$ represents the cumulative microscopic noise
for this interval. For the next interval $[1,2)$ on the fast scale
(corresponding to $[\epsilon,2\epsilon)$ on the slow scale) we can
repeat the process and obtain $\zeta_1=\int_1^2\xi(x)dx$. In this way
we obtain a sequence of independent normally distributed random
numbers $(\zeta_k)=(\zeta_0,\zeta_1,...)$.  On the slow scale, the
random number $\zeta_k$ characterizes the noise in the $k$-th interval
$[k\epsilon,(k+1)\epsilon)$. Hence, we can view the sequence of random
numbers $(\zeta_k)$ on the slow scale as a discretization of a white
noise process $\Xi=\Xi(x_1)$ with $\langle \Xi(x_1)\Xi(x_1')\rangle =
\epsilon \delta(x_1-x_1')$.  In this way we obtain the stochastic
short pulse equation with coarse-grained noise as
\begin{equation} \label{stochastic_SPE}
  -2(M_0)_{x_1 \phi} = \left(\bar\chi_0+ \nu \Xi(x_1)\right)M_0
                       + \chi_3(M_0 ^3)_{\phi \phi} 
\end{equation}
One important difference between the deterministic and the stochastic
case is that, if we introduce (\ref{stochastic_SPE}) back in
(\ref{equation_for_M1}), we obtain a dependence of $M_1$ on the fast
variable $x_0$ that is not present in the deterministic case. This is
especially important when considering the next order
${\mathcal{O}}(\epsilon^2)$ in the asymptotic expansion. In this
work, however, we only consider terms up to the order
${\mathcal{O}}(\epsilon)$.

Note that we obtain the strength of the slow-scale noise $\Xi$
explicitly and that both the microscopic strength of the noise (given
by $\nu$) and the expansion coefficient $\epsilon$ influence the
strength of the coarse-grained noise. The reason for the fact that the
coarse-graining of the noise takes a rather simple form lies in the
fact that the randomness is introduced in the linear dispersive term
and that the leading order evolution is trivial as $M_0$ is
independent of $x_0$. In cases where the leading order solution is
more complicated, its structure will influence the strength of the
coarse-grained noise \cite{schaefer-moore:2007}.

Aside from fluctuations in the dispersion, other parameters might vary stochastically as well. In deriving the short pulse equation from Maxwell's equations, an approximate fit of the linear susceptibility of the form 
\begin{equation}\label{linearSuscepApprox}
\hat{\chi}^{(1)}(\lambda)\approx\hat{\chi}_0^{(1)} - \hat{\chi}_2^{(1)} \lambda^2
\end{equation}
in Fourier domain \cite{schaefer-wayne:2004,Levent2011} was made. This susceptibility approximation is found by fitting the experimental data for light propagation in silica fibers in the infrared regime. The first stochastic generalization of the SPE treated above is derived by considering the fluctuations in the $\lambda$-dependent term. However, one can argue that the fluctuations may come from the $\hat{\chi}_0^{(1)}$ term in (\ref{linearSuscepApprox}) rather than the $\hat{\chi}_2^{(1)}$ term, corresponding to stochastic variations of the phase velocity. This treatment will lead to a different stochastic wave equation:   
\begin{equation} \label{maxwell_stochastic2}
u_{xx}- \left[1+\nu\xi(x)\right] u_{tt} = \chi_0 u +\chi_3(u^3)_{tt}\;.
\end{equation}
As before, a multiple scale expansion, see (\ref{uExpansion}) and (\ref{variablesExpansion}), leads to a different form of the stochastic SPE. As the calculations are similar to the previous case of random dispersion, we just give the result:
\begin{equation} \label{stochastic_SPE_2}
-2(M_0)_{x_1 \phi} = \chi_0 M_0
                       + \chi_3 (M_0 ^3)_{\phi \phi} + \left(\frac{\nu}{\epsilon^2} \Xi(x_1)\right) (M_0)_{\phi \phi}
\end{equation}

Finally, we will introduce the noise in the nonlinearity as well, which will result in a stochastic wave equation in the form of
\begin{equation} \label{maxwell_stochastic3}
u_{xx}-u_{tt} = \chi_0 u + \left[\bar\chi_3 +\nu\xi(x)\right] (u^3)_{tt}\;.
\end{equation}
The corresponding stochastic SPE takes the form   
\begin{equation} \label{stochastic_SPE_3}
-2(M_0)_{x_1 \phi} = \chi_0 M_0
                       + \left(\bar\chi_3+ \nu \Xi(x_1)\right)(M_0 ^3)_{\phi \phi} \;. 
\end{equation}
If one sets $\nu =0$ in any of these stochastic generalizations of the SPE, one will obtain the deterministic short pulse equation (\ref{spe_sw}).

\section{Numerical comparison} \label{sec:numerics}

\subsection{Pathwise comparison}

Again, we expect (\ref{stochastic_SPE}) to be a decent approximation
of the nonlinear wave equation (\ref{maxwell_1d}) with a stochastic
linear susceptibility given by (\ref{stochastic_chi}) for propagation
distances that are of the order of at least
${\mathcal{O}}(1/\epsilon)$. Since we obtained a very good agreement
up to distances of ${\mathcal{O}}(1/\epsilon^2)$ in the deterministic
case, we will consider this as our maximum propagation distance for
the stochastic case as well. 

In order to evaluate the quality of the approximation numerically, we
can look at a {\em path-wise} comparison. From the coarse-graining
procedure described above, we obtain for any particular realization
$\xi_p$ of the micro-scale noise $\xi$ on the fast scale a macroscopic
noise $\Xi_p$ on the slow scale. Then, we can compare the solution of
the stochastic SPE (\ref{stochastic_SPE}) with the solution of the
nonlinear wave equation for this particular realization of the
noise. Again we choose to look at the differences of the solution of
the stochastic PDEs and the deterministic soliton. For the simulation
of the nonlinear wave equation, we subtract the error that exists in
the deterministic system, presented in figure
\ref{fig:NormDifference}. As a result, we obtain two stochastic
processes - one given by the difference of the solution of the
stochastic SPE and the deterministic soliton and the other given by
the difference of the solution of the stochastic nonlinear wave equation and the
 deterministic solution to the wave equation (with the SPE soliton as initial condition). 

\begin{figure}[htb]
\centering
\scalebox{0.4}{\includegraphics{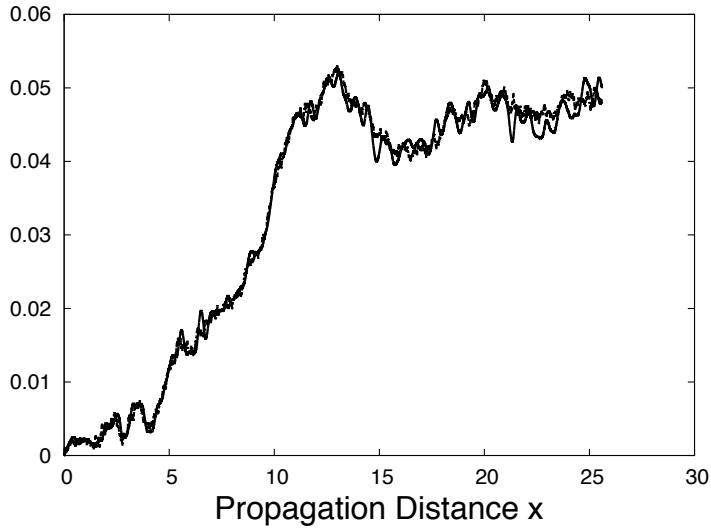}}
\hfill
\caption{Stochastic dispersion: Evolution of the stochastic difference of the SPE soliton and
  the solution of the nonlinear wave equation (\ref{maxwell_1d}). The
  solid line shows the evolution of the $L^{\infty}$-norm of the difference
  generated by the solution of the stochastic nonlinear wave equation
  and the dashed line shows the corresponding difference for the corresponding 
  evolution in the stochastic short pulse equation. The strength parameter 
  $\nu$ of the microscopic noise was 0.125.}
\label{fig:Disp}
\end{figure}

The result
of the corresponding numerical simulations is shown in figure
\ref{fig:Disp} and shows excellent agreement of the
approximation and the solution of the original equation.

%

The next figure \ref{fig:GV} shows the same comparison for the case of randomness 
in the phase velocity. In this case, we compare the solutions to (\ref{stochastic_SPE_2})
to the solutions of (\ref{maxwell_1d}) for the case of random variations of
$\hat\chi^{(1)}_0$. 

\begin{figure}[htb]
\centering
\scalebox{0.4}{\includegraphics{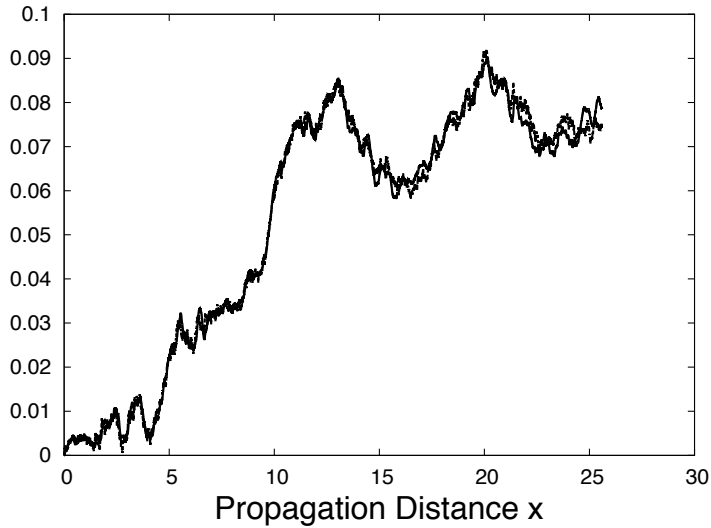}}
\hfill
\caption{Stochastic phase velocity: Evolution of the stochastic difference of the SPE soliton and
  the solution of the nonlinear wave equation (\ref{maxwell_1d}) for the case of random phase velocity. The
  solid line shows the evolution of the $L^{\infty}$-norm of the difference
  generated by the solution of the stochastic nonlinear wave equation
  and the dashed line shows the corresponding difference for the corresponding 
  evolution in the stochastic short pulse equation. The strength parameter 
  $\nu$ of the microscopic noise was 0.005. Recall that $\epsilon = 0.2$ such that
the ratio $\nu/\epsilon^2$ is still fairly small.}
\label{fig:GV}
\end{figure}

\begin{figure}[htb]
\centering
\scalebox{0.4}{\includegraphics{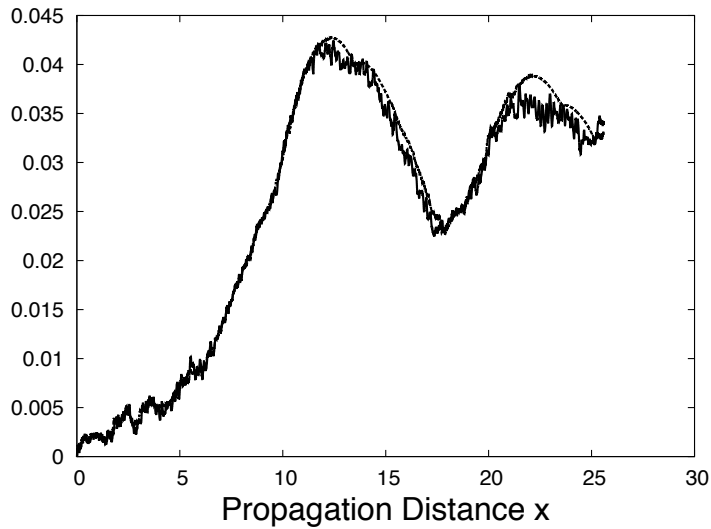}}
\hfill
\caption{Stochastic phase velocity: Evolution of the stochastic difference of the SPE soliton and
  the solution of the nonlinear wave equation (\ref{maxwell_1d}) for the case of random phase velocity. The
  solid line shows the evolution of the $L^{\infty}$-norm of the difference
  generated by the solution of the stochastic nonlinear wave equation
  and the dashed line shows the corresponding difference for the corresponding 
  evolution in the stochastic short pulse equation. The strength parameter 
  $\nu$ of the microscopic noise was 0.05.}
\label{fig:NL}
\end{figure}

The third case is the comparison for the case of random fluctuation of the nonlinear
coefficient. Figure \ref{fig:NL} compares the solution of the stochastic short pulse equation
with the solution to stochastic Maxwell's equations.

Hence, for all three cases, we have numerical evidence that the coarse-grained stochastic generalizations of the SPE capture correctly the impact of the noise on the solitons when 
propagating in the stochastic nonlinear wave equation.

\subsection{Probability distribution}

This pathwise correspondence indicates that we can also expect a
decent agreement in the prediction of the related probability
distributions.  To check this, we can plot the histograms of the
deviations of the solutions to the stochastic PDEs from the
deterministic evolution of the SPE solitons. The following figure
\ref{fig:Statistics} was obtained by creating a probability
distributions from 10000 realizations of the stochastic short pulse
equation and stochastic Maxwell's equation for the case of 
stochastic variations of the dispersion.

\begin{figure}[htb]
\centering
\scalebox{0.4}{\includegraphics{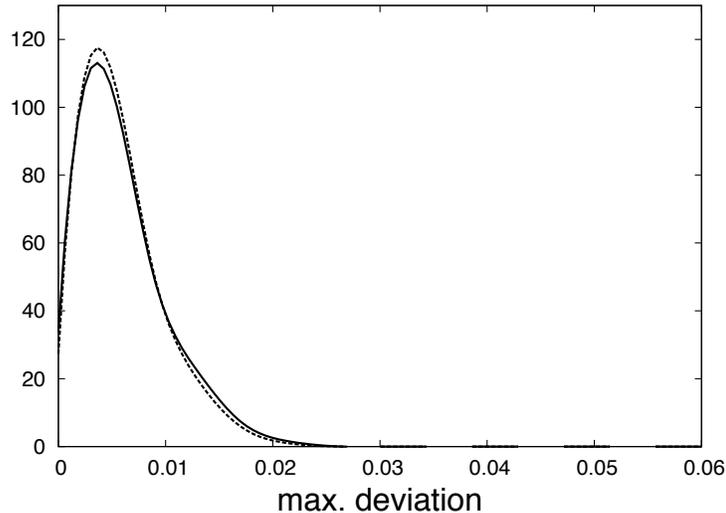}}
\hfill
\caption{Probability distributions of the stochastic difference of the
  SPE soliton and the solution of the nonlinear wave equation
  (\ref{maxwell_1d}) at the propagation distance $x = 25.625$. The
  $x$-axis shows the $L^{\infty}$-norm of the difference generated by
  the solution of the stochastic PDEs in comparison to the
  deterministic solution (solid line: Maxwell, dashes: SPE).  
In this way, the solution of the nonlinear
  wave equation has been corrected for the deterministic error. The strength of the
noise was $\nu = 0.05$.}
\label{fig:Statistics}
\end{figure}

Although we see a small difference of the SPE and stochastic Maxwell's
equations for small deviations, we obtain overall an excellent agreement between
the approximation given by the stochastic short pulse equation and the
stochastic nonlinear wave equation. In particular, the shape of the probability
distribution is correctly predicted by the SPE approximation and large deviations
from the mean are correctly accounted for as well.

\section{Conclusion}

In this paper we have derived stochastic generalizations of the short
pulse equation and shown explicitly how the noise processes governing
the parameters of the SPE can be constructed from the noise present in
the underlying nonlinear wave equation. Moreover, we have shown that
SPE solitons persist in the nonlinear wave equation with stochastic variations
present and that they exhibit only
small changes over the propagation distances under consideration. The
changes arising from the random part of the susceptibility can be
correctly characterized by the stochastic short pulse equation.

\section*{Acknowledgments}
This work was partially supported by National Science
Foundation through the grants DMS-0807396 and DMS-1108780 and through the grant
of the PSC-CUNY research foundation PSCREG-39-709. The authors gratefully acknowledge the support of the
CUNY High Performance Computing Facility and the Center for
Interdisciplinary Applied Mathematics and Computational Sciences. The
authors are grateful to D. Trubatch for many valuable discussions.

\bibliography{masterPaper}

\end{document}